\documentstyle[epsf]{article}

\begin {document}

\begin{center}
{\Large\bf Dynamical Properties of one dimensional Mott Insulators}\vskip .5cm

{\large Davide Controzzi$^{(a)}$, Fabian H.L. Essler$^{(b)}$ and
Alexei M. Tsvelik$^{(a)}$}\\ 
\vskip .5cm
{$^{(a)}$ Department of Physics, University of Oxford, 1 Keble Road, Oxford OX1 3NP, UK}\\
{$^{(b)}$ Department of Physics, Warwick University, Coventry CV4 7AL, UK}
\end{center}

\begin{abstract}
\par
At low energies the charge sector of one dimensional Mott insulators
can be described in terms of a quantum Sine-Gordon model (SGM). 
Using exact results derived from integrability it is possible to
determine dynamical properties like the frequency dependent optical
conductivity. We compare the exact results to perturbation theory and
renormalisation group calculations. We also discuss the application of
our results to experiments on quasi-1D organic conductors.
\end{abstract}

{Lecture given by FHLE at the NATO ASI/EC summer school {\sl ``New Theoretical
Approaches to Strongly Correlated Systems''}, Sir Isaac Newton Insitute
for Mathematical Sciences, Cambridge April 2000}

\section{1D Mott insulators}
The Mott metal-insulator transition is a paradigm for the importance
of electron-electron interactions in condensed matter systems. It
occurs in a variety of actual materials and has attracted much
attention over the last fifty years \cite{mott}. The underlying
mechanism that drives the transition is by now well understood, but
details on e.g. transport properties remain largely unknown in $D=2$
and $D=3$ due to the lack of nonperturbative methods for treating
strongly correlated electron systems. The situation is more fortunate
in two cases: $D=\infty$, where much progress has been made in recent
years \cite{dinfty} and $D=1$, where nonperturbative methods permit
essentially a full solution of the problem. The 1D case is the one we
will be concerned with here.
A full characterization of the Mott insulating phase requires the
knowledge of dynamical correlation functions. The frequency dependent
optical conductivity $\sigma(\omega)$ is one of the most important
examples from an experimental point of view. The behaviour of
$\sigma(\omega)$ in the metallic regime is easily understood in terms
of the Tomonaga-Luttinger theory \cite{GNT,LL}. The situation in the Mott
insulating phase is much more complicated due to the spectral gap that
is dynamically generated by the electron-electron interactions. Here 
$\sigma(\omega)$ has until now only been studied by perturbative
methods \cite{gia1,giamarchi}, which break down in the most interesting
regime of frequencies close to the optical gap. In these proceedings
we use methods of integrable quantum field theory to determine
$\sigma(\omega)$ in 1D Mott insulators for all frequencies much
smaller than the bandwidth, which is the large scale in the field
theory approach to the problem. Some of the results presented here have
already appeared in \cite{cet}.

The paradigm of a 1D Mott insulator is the Hubbard model
\begin{equation}
H = -t \sum_{l;\sigma} \left( c_{l,\sigma}^\dagger c_{l+1,\sigma} + {\rm h.c.}\right) +
U \sum_{l} {n}_{l,\uparrow}{n}_{l,\downarrow} \, .
\label{Hamiltonian}
\end{equation}
Here $c_{l,\sigma}$ are fermionic annihilation operators of spin
$\sigma=\uparrow,\downarrow$ at site $l$ of a one-dimensional chain
and $n_{l,\sigma}= c^\dagger_{l,\sigma}c_{l,\sigma}$. The 1D Hubbard
model is solvable by the Bethe Ansatz \cite{LiebWu} and the exact
solution establishes the presence of a Mott transition at half-filling
(one electron per site) and $U=0$. In other words, for a half-filled
band the model is insulating for any nonzero value of the on-site
repulsion $U$, whereas it is metallic for a less than half-filled band.
Many properties have been determined exactly \cite{Korepinbuch}, but
the asymptotic behaviour of dynamical correlation functions are at
present known only in the metallic \cite{FK} and the gas phase
\cite{GK}.
In order to make progress in the insulating phase and to clearly expose
the mechanism underlying the transition it is very useful to consider
the scaling limit of the half-filled Hubbard model. This field theory limit
corresponds to weak coupling and can be obtained directly
from the exact spectrum and scattering matrix of the lattice model
\cite{smat,ezer}. It is defined by 
\begin{equation}
t\to\infty\ ,\quad U/t\to 0\ ,\quad
M= \frac{8t}{\pi}\sqrt{\frac{U}{4t}}\exp\left(-2\pi t/U\right){\rm fixed}.
\end{equation}
and has been studied by many authors \cite{scaling}.
On an operator level the field theory can be constructed along
the lines of e.g. chapter 15 of \cite{GNT}. One starts by splitting
the electron operators into fast and slow components 
\begin{equation}
c_{l,\sigma}\longrightarrow \sqrt{a_0} \left[\exp(ik_F x)\ R_\sigma(x)+\exp(-ik_F x)\
L_\sigma(x)\right] .
\label{cpsi}
\end{equation}
Here $k_F$ is the Fermi momentum (it is $\pi/2$ for the half-filled
band), $R_\sigma$ and $L_\sigma$ are right and left moving electron
fields and $x=l a_0$, where $a_0$ is the lattice spacing. Inserting
this prescription into the Hamiltonian (\ref{Hamiltonian}) one obtains
\begin{eqnarray}
{\cal H}&=& \sum_{\sigma} v_F \int dx\left[L_\sigma\ i\partial_x
L^\dagger_\sigma - R_\sigma\ i \partial_x R^\dagger_\sigma\right]
+g\int dx\left[ {\bf I}\cdot {\bf\bar{I}} - {\bf J}\cdot
{\bf\bar{J}}\right]\nonumber\\ 
&&\qquad+\frac{g}{6}\int dx\left[:{\bf I}\cdot{\bf I}: + :{\bf
\bar{I}}\cdot {\bf\bar{I}}:
-:{\bf J}\cdot {\bf J}: - :{\bf \bar{J}}\cdot {\bf\bar{J}}: \right]\ ,
\label{hamil1}
\end{eqnarray}
where $v_F=2ta_0$ is the Fermi velocity and $g=2U a_0$. Here $\bf J$
and $\bf I$ are the chiral components of SU(2) spin and pseudospin
currents 
\begin{eqnarray}
I^3&=&\frac{1}{2}\sum_\sigma :L^\dagger_\sigma L_\sigma :\ ,\quad
I^+=L^\dagger_\uparrow L^\dagger_\downarrow\ ,\nonumber\\
\bar{I}^3&=&\frac{1}{2}\sum_\sigma :R^\dagger_\sigma R_\sigma :\ ,\quad
\bar{I}^+=R^\dagger_\uparrow R^\dagger_\downarrow\ ,\nonumber\\
{J}^3&=&\frac{1}{2}\left(L^\dagger_\uparrow L_\uparrow
-L^\dagger_\downarrow L_\downarrow\right)\ ,\quad
{J}^+=L^\dagger_\uparrow L_\downarrow\ ,\nonumber\\
\bar{J}^3&=&\frac{1}{2}\left(R^\dagger_\uparrow R_\uparrow
-R^\dagger_\downarrow R_\downarrow\right)\ ,\quad
\bar{J}^+=R^\dagger_\uparrow R_\downarrow\ .
\end{eqnarray}
Note that the Hamiltonian (\ref{hamil1}) displays the required SO(4)
symmetry \cite{so4} of the half-filled Hubbard model. 
By employing the Sugawara construction, the Hamiltonian (\ref{hamil1})
can now be split into two parts, corresponding to the spin and charge
sectors respectively \cite{GNT}
\begin{eqnarray}
{\cal H}&=& {\cal H}_c+{\cal H}_s\ ,\nonumber\\
{\cal H}_c&=& \frac{2\pi v_c}{3}\int dx\left[
:{\bf I}\cdot {\bf I}:+:\bar{\bf I}\cdot \bar{\bf I}:\right]
+g\int dx\ {\bf I}\cdot \bar{\bf I}\ ,\nonumber\\
{\cal H}_s&=& \frac{2\pi v_s}{3}\int dx\left[
:{\bf J}\cdot {\bf J}:+:\bar{\bf J}\cdot \bar{\bf J}:\right]
-g\int dx\ {\bf J}\cdot \bar{\bf J}\ .
\label{su2thi}
\end{eqnarray}
Here $v_s=v_F-Ua_0/2\pi$ and $v_c=v_F+Ua_0/2\pi$.
Apart from the (marginally) irrelevant current-current interaction in
the spin sector and the difference in spin and charge velocities, the
Hamiltonian (\ref{su2thi}) is identical to the one of the SU(2)
Thirring model \cite{su2th}. The latter is integrable
\cite{su2th2,jnw} and using its exact solution it is possible to
determine dynamical correlation functions via the formfactor bootstrap
approach. For example the optical conductivity was determined in this
way and compared to numerical dynamical density matrix renormalisation
group computations on the Hubbard model in \cite{hubb}. Here we 
consider the optical conductivity for a more general model. We note
that the optical conductivity is rather special in that for all cases
considered here the electric current operator couples only to the
charge sector, which greatly simplifies all calculations. 
Let us now consider an extended Hubbard model
\begin{equation}
H_{\rm ext} = -t \sum_{l;\sigma} \left( c_{l,\sigma}^\dagger
c_{l+1,\sigma} + {\rm h.c.}\right) + U \sum_{l}
{n}_{l,\uparrow}{n}_{l,\downarrow}
+ V \sum_{l} {n}_{l}{n}_{l+1} \ ,
\label{Hext}
\end{equation}
where $n_l=n_{l,\uparrow}+n_{l,\downarrow}$. By repeating the above
analysis we find that the scaling limit takes the form

\begin{eqnarray}
{\cal H}_{\rm ext}&=& {\cal H}_c^\prime+{\cal H}_s^\prime\ ,\nonumber\\
{\cal H}_c^\prime&=& \frac{2\pi v_c^\prime}{3}\int dx\left[
:{\bf I}\cdot {\bf I}:+:\bar{\bf I}\cdot \bar{\bf I}:\right]
+\int dx\left[g_\perp
\left(I^+\bar{I}^-+I^-\bar{I}^+\right)+g_\parallel\ I^z \bar{I}^z \right]
 ,\nonumber\\
{\cal H}_s^\prime&=& \frac{2\pi v_s}{3}\int dx\left[
:{\bf J}\cdot {\bf J}:+:\bar{\bf J}\cdot \bar{\bf J}:\right]
-2g_\perp\int dx\ {\bf J} \cdot \bar{\bf J}.
\label{Hext2}
\end{eqnarray}
where 
\begin{equation}
g_\perp=(U-2V)a_0\ ,\quad g_\parallel=(2U+12V)a_0\ ,\quad 
v_c^\prime= v_F+\frac{(U+4V)a_0}{2\pi}.
\end{equation}
Clearly spin-charge separation still holds, so that the two parts of
the Hamiltonian (\ref{Hext2}) can be bosonized separately. Here we are
interested in the charge sector only but note in passing that the
spin sector is gapless if $U>2V$ and gapped otherwise. Applying the
standard bosonization rules to ${\cal H}_c^\prime$ in (\ref{Hext2}) one
arrives at the Hamiltonian of the SGM (\ref{sg}) \cite{GNT}.

The electric current operator of the lattice models
(\ref{Hamiltonian}), (\ref{Hext}) is given by
\begin{equation}
{j}=\frac{-iet}{\hbar}\sum_{j,\sigma}
\left[c^\dagger_{j,\sigma}c_{j+1,\sigma}-
c^\dagger_{j+1,\sigma}c_{j,\sigma}\right] ,
\end{equation}
and does not commute with the above Hamiltonians. Using
(\ref{cpsi}) this becomes
\begin{equation}
{j}= \frac{4et}{\hbar}\int dx\left[I^3(x)-\bar{I}^3(x)\right]
\label{elcu}
\end{equation}
in the field theory limit.
From now on we drop the factor $e t/\hbar$ which simply fixes the
units in which we measure the current. 

\section{The Sine-Gordon model}

The low energy physics of the charge sector of a general, pure,
one-dimensional Mott insulator is described by the SGM
as we have seen in the previous section for some specific examples. 
The action is given by
\begin{equation}
\label{sg}
{\cal S}_{SG} = \int \; d^2 x \left \{ \frac{1}{16\pi}(\partial_\nu
\phi )^2 
-2 \mu \cos(\beta \phi) \right \}\ .
\end{equation}
Here we have chosen the normalization of the Bose field following
\cite{abz,lukyanov} by specifying the short-distance behaviour of the
two-point function as
\begin{equation}
\langle e^{i\alpha\phi(x)}\ e^{-i\alpha\phi(y)}\rangle\longrightarrow
|x-y|^{-4\alpha^2}\quad {\rm as}\ |x-y|\to 0.
\end{equation}

The SGM posesses a conserved (topological) charge
\begin{equation}
Q=\int_{-\infty}^\infty \; j^0 \ dx=-\frac{\beta}{2\pi}
\int _{-\infty}^\infty \frac{\partial
\phi}{\partial x}\ dx\ ,
\end{equation}
where 
\begin{equation}
\label{j}
j^\mu=-\frac{\beta}{2\pi}\epsilon^{\mu \nu}\partial _\nu \phi
\end{equation}
is the Noether current \footnote{It is normalized such that solitons and
antisolitons have charges $\pm 1$.}. The electric current (\ref{elcu})
is proportional to the Noether current $j^1$
\begin{equation}
{ j}=\sqrt{A}\ \partial_0\phi\ ,
\label{elcurr}
\end{equation}
where ${A}$ is some nonuniversal constant
\footnote{We assume that $A$ is nonuniversal because the electric
current is not a conserved quantity for the lattice model.}. The SGM
is integrable and has been studied in great detail over the last 25
years \cite{SG1,SG,SG2}. Let us review some results obtained from the
exact solution \cite{SG} that we will need in the following. First of
all some of the results obtained in the repulsive regime $\beta^2 >1/2$
appear to depend on the regularization scheme \cite{SG2,jnw} employed
to deal with the UV divergences. Here we follow the method of \cite{jnw},
which is very natural from a field-theory point of view.

The spectrum of the SGM depends on the value of the coupling constant
$\beta^2$ or, alternatively, on 
\begin{equation}
\xi=\frac{\beta^2}{1-\beta^2}.
\end{equation}
For $\beta^2<1$ the cosine term is relevant in the renormalization group
(RG) sense and dynamically generates a spectral gap $M $ in the
excitation spectrum. In the repulsive regime $1<\xi <\infty$ the
spectrum contains only charged particles of charge $Q=\pm 1$, which
are called {\sl solitons} and {\sl antisolitons}. In this regime the
spectral gap is related to the ``optical gap'' $\Delta$ i.e. the gap
seen in the optical absorption spectra by $\Delta=2 M$. At the  
so-called ``Luther-Emery'' \cite{LuE} (LE) point $\xi = 1$ the SGM is equivalent 
to a free massive Dirac fermion. In this limit the solitons become
non-interacting particles and as we will see the Mott insulator turns
into a conventional band insulator.
In the  limit $\xi \rightarrow  \infty$ the sine-Gordon model 
acquires an SU(2) symmetry and describes the charge sector of the
Hubbard model at half-filling in the limit of weak interactions as
discussed above. 
In the attractive regime $0<\xi <1$ excitonic soliton-antisoliton
bound states are formed and the spectrum becomes more complicated.
Here we constrain ourselves to the repulsive regime and refer to
\cite{EGJ} for results on dynamical correlation functions in extended
Hubbard models that correspond to the attractive regime in the SGM.

As usual in a theory with relativistic dispersion $e(p)=\sqrt{p^2+M
^2}$ it is useful to parametrize the spectrum in terms of a rapidity
variable $\theta$ defined by
\begin{equation}
\label{ep} 
p=M\sinh\theta, \; e = M\cosh\theta.
\end{equation}
Let us distinguish solitons and antisolitons by an index
$\varepsilon = \pm$.
The exact 2-particle soliton-antisoliton scattering matrix is then given by
\cite{smatrix:sg}
\begin{eqnarray}
S_{+,+}^{+,+}(\theta)&=&
S_{-,-}^{-,-}(\theta)=
S_0(\theta),\nonumber\\
S_{+,-}^{+,-}(\theta)&=&
S_{-,+}^{-,+}(\theta)=
-\frac{\sinh\frac{\theta}{\xi}}{\sinh\frac{1}{\xi}(\theta-\pi i)}
S_0(\theta),\nonumber\\
S^{+,-}_{-,+}(\theta)&=&
S^{-,+}_{+,-}(\theta)=
-\frac{\sinh\frac{\pi i}{\xi}}{\sinh\frac{1}{\xi}(\theta-\pi i)}
S_0(\theta),\nonumber\\
S_{a,b}^{a',b'}(\theta)&=&0 \quad a'+b'\neq a+b,\nonumber\\
S_0(\theta)&=&-\exp \left \{-i \int_0 ^\infty \frac{\sin(\theta t/\pi) \;
\sinh \left (
\frac{1-\xi}{2}t \right )}{t \; \cosh \left (\frac{t}{2} \right
) \; \sinh \left ( \frac{\xi t}{2} \right ) } dt \right \}.
\label{smat}
\end{eqnarray}
The two-particle S-matrix (\ref{smat}) completely specifies all
scattering processes in the SGM as multi-particle scattering
is purely elastic and factorizes into two-particle processes.
A convenient formalism for the description of a dilute gas of
particles with factorizable scattering is obtained in terms of
the Zamolodchikov-Faddeev (ZF) algebra. The ZF algebra can be
considered to be the logical extension of the algebra of creation and
annihilation operators for free fermion or bosons to the case of
interacting particles with factorizable scattering.
The ZF algebra is usually introduced formally based on the knowledge
of the exact spectrum and scattering matrix, which for the SGM was
obtained in \cite{zamo,karo}. For the SGM in the repulsive regime the
ZF operators (and their hermitian conjugates) thus satisfy the
following algebra 
\begin{eqnarray}
{Z}^{\varepsilon_1}(\theta_1){Z}^{\varepsilon_2}(\theta_2) &=& S^{\varepsilon_1,\varepsilon_2}_
{\varepsilon_1',\varepsilon_2'}(\theta_1 -
\theta_2){Z}^{\varepsilon_2'}(\theta_2){Z}^{\varepsilon_1'}(\theta_1)\ ,
\nonumber\\
{Z}_{\varepsilon_1}^\dagger(\theta_1)Z_{\varepsilon_2}^\dagger(\theta_2) &=&
Z_{\varepsilon_2'}^\dagger(\theta_2){ Z}_{\varepsilon_1'}^\dagger
(\theta_1)S_{\varepsilon_1,\varepsilon_2}^{\varepsilon_1',\varepsilon_2'}(\theta_1 -
\theta_2) , \nonumber\\
Z^{\varepsilon_1}(\theta_1)Z_{\varepsilon_2}^\dagger(\theta_2) &=&Z_{\varepsilon_2'}
^\dagger(\theta_2)
S_{\varepsilon_2,\varepsilon_1'}^{\varepsilon_2',\varepsilon_1}(\theta_2-\theta_1)Z^{\varepsilon_1'}
(\theta_1) 
+(2 \pi) \delta_{\varepsilon_2}^{\varepsilon_1} 
\delta (\theta_1-\theta_2), \quad
\varepsilon_1,\varepsilon_2
=\pm\frac{1}{2}.
\label{fz1}
\end{eqnarray}
Here the two-particle scattering matrices $S^{\varepsilon_1,\varepsilon_2}_{\varepsilon_1',\varepsilon_2'} 
(\theta)$ are defined in Eq.(\ref{smat}) and $\varepsilon_j=\pm$.
The factor $2\pi$ in the last equation stems from the normalization of
the single particle asymptotic states (cf Eq.(\ref{identity})).

Using the ZF generators a Fock space of states can be constructed as
follows. The vacuum is defined by
\begin{equation}
Z_{\varepsilon}(\theta) |0\rangle=0 \ .
\end{equation}
Multiparticle states are then obtained by acting with strings of
creation operators $Z_\varepsilon^\dagger(\theta)$ on the vacuum
\begin{equation}
|\theta_n\ldots\theta_1\rangle_{\varepsilon_n\ldots\varepsilon_1} = 
Z^\dagger_{\varepsilon_n}(\theta_n)\ldots Z^\dagger_{\varepsilon_1}(\theta_1)|0\rangle .
\label{states}
\end{equation} 
We note that (\ref{fz1}) together with
(\ref{states}) implies that states with different orderings of two
rapidities and indices $\varepsilon_i$ are related in the following way
\begin{equation}
|\theta_n\ldots\theta_k\theta_{k+1}\ldots\theta_1
\rangle_{\varepsilon_n\ldots\varepsilon_k\varepsilon_{k+1}\ldots\varepsilon_1} = 
S_{\varepsilon_k,\varepsilon_{k+1}}^{\varepsilon'_{k},\varepsilon'_{k+1}}(\theta_k-\theta_{k+1})
|\theta_n\ldots\theta_{k+1}\theta_k\ldots\theta_1
\rangle_{\varepsilon_n\ldots\varepsilon'_{k+1}\varepsilon'_k\ldots\varepsilon_1}.
\label{order}
\end{equation}
The resolution of the identity is given by
\begin{equation}
1\!\!1=\sum_{n=0}^\infty\sum_{\varepsilon_i}\int_{\infty}^{\infty}
\frac{d\theta_1\ldots d\theta_n}{(2\pi)^nn!}
|\theta_n\ldots\theta_1\rangle_{\varepsilon_n\ldots\varepsilon_1}
{}^{\varepsilon_1\ldots\varepsilon_n}\langle\theta_1\ldots\theta_n|\ .
\label{identity}
\end{equation}

\section{Spectral representation of the optical conductivity}

An efficient method for the computation of correlation functions 
in integrable, massive quantum field theories is given by the form
factor approach. This approach is based on the spectral
representation, that expresses correlation functions in terms of an
infinite series over multi-particles states. The two-point correlation
function of some operator ${\cal O}$ can be written as 
\begin{equation}
\label{corr}
\langle {\cal O}(x,t){\cal O}^\dagger(0,0)\rangle
=\sum_{n=0}^\infty\sum_{\varepsilon_i}\int
\frac{d\theta_1\ldots d\theta_n}{(2\pi)^nn!}
\exp\left({i\sum_{j=1}^n p_jx-e_j t}\right)
|\langle 0| {\cal O}(0,0)|\theta_n\ldots\theta_1
\rangle_{\varepsilon_n\ldots\varepsilon_1}|^2, 
\end{equation}
where 
$p_j$ and $e_j$ have the form (\ref{ep})
\begin{equation}
p_j=M\sinh \theta_j, \; e_j=M\cosh \theta_j,
\end{equation}
and 
\begin{equation}
\label{formf}
f^{\cal O}(\theta_1\ldots\theta_n)_{\varepsilon_1\ldots\varepsilon_n}\equiv
\langle 0| {\cal
O}(0,0)|\theta_n\ldots\theta_1\rangle_{\varepsilon_n\ldots\varepsilon_1} 
\end{equation}
are the form factors (FF).
They can be calculated using a set of ``axioms'' specifying their
analytical properties (see \cite{smirnov,karow,lukyanov,ffother} and
H. Saleur's contribution to this volume).  
From a physical point of view we will be mainly interested in
Fourier transforms of retarded two point functions. Their form factor
expansions have the form
\begin{eqnarray}
\label{expansion1}
&&\chi^{\cal O}(\omega,q)= i \int_{-\infty}^\infty d x\int_0^\infty d t
\ e^{i (\omega+i\varepsilon)t-iqx} \langle
[{\cal O}(x,t) , {\cal O}^\dagger(0,0)]\rangle \nonumber\\
&&= -2\pi\sum_{n=0}^\infty\sum_{\varepsilon_i}\int
\frac{d\theta_1\ldots d\theta_n}{(2\pi)^nn!}
|f^{\cal O}(\theta_1\ldots\theta_n)_{\varepsilon_1\ldots\varepsilon_n}|^2 \nonumber\\
&&\times 
\left\lbrace\frac{\delta(q-M\sum_j\sinh\theta_j)}{\omega - M \sum_j
\cosh\theta_j +i\epsilon}-
\frac{\delta(q+M \sum_j\sinh\theta_j)}{\omega + M \sum_j
\cosh\theta_j +i\epsilon}\right\rbrace .\nonumber\\
\end{eqnarray}

We will be interested in the optical conductivity which is related to
the imaginary part of the retarded current-current correlation function, 
$\chi^{j}(\omega,q)$, by
\begin{equation}
\label{sigma}
\sigma(\omega >0) = {\rm Im}
\left \{ \chi^{j}(\omega,q=0) \right \} /\omega .
\end{equation}
Here the electric current operator $j$ was defined in (\ref{elcurr}).
Using (\ref{expansion1}) we obtain the following representation for the
optical conductivity
\begin{eqnarray}
\sigma(\omega)&=& \frac{2 \pi^2}{\omega}
\sum_{n}\sum_{\varepsilon_i}\int 
\frac{ d\theta_1\ldots d\theta_n}{(2\pi)^n n!} \left |
f^j(\theta_1\ldots\theta_n)_{\varepsilon_1\ldots\varepsilon_n}
\right | ^2 \nonumber\\
 &\times&\delta(M \sum_k\sinh\theta_k)\delta(\omega - M \sum_k
\cosh\theta_k) \nonumber\\
&=&\sigma_2(\omega)+\sigma_4(\omega)+ ...
\label{expansion2}
\end{eqnarray}
Here 
$
\label{ff}
f^{j}(\theta_1\ldots\theta_n)_{\varepsilon_1\ldots\varepsilon_n}\equiv
\langle 0| j(0,0)|\theta_n\ldots\theta_1\rangle_{\varepsilon_n\ldots\varepsilon_1}
$
are the form factors of the electric current operator (\ref{elcurr}), 
$\sigma_2(\omega)$ and $\sigma_4(\omega)$ represent the
contributions from 2 and 4-particle processes and the dots
indicate processes involving higher numbers of (anti)solitons. 
We note that as a consequence of charge conjugation symmetry only 
intermediate states with an even number of particles contribute 
to this correlation function \cite{smirnov}. From (\ref{expansion2})
it is easy to see that only 2-particle processes contribute up to
energies $\omega=4 M$, only 2 and 4-particle processes up to $\omega=6
M$ and so on. It has been previously observed for several models that
the FF series converges much more rapidly than expected on the basis
of such considerations \cite{oldff1,oldff2,o3,delf-muss}. This may be
understood in terms of phase space arguments \cite{oldff1,mussardo.school}. 

The $n$-particle form factors (\ref{formf}) for the current operator
$j^\mu$ in the SGM have been determined in \cite{smirnov} and can be
used to calculate the first few terms in the expansion
(\ref{expansion2}). The two particle FF is given 
by
\begin{equation}
f^j(\theta_1,\theta_2)_{+-}=-f^j(\theta_1,\theta_2)_{-+}
=\frac{4\pi^2 M\sqrt{A}}{\beta} \xi d 
\frac{\cosh\frac{\theta_1+\theta_2}{2}}{\cosh{ \left ( \frac{(\theta_1-
\theta_2) +i \pi}{2 \xi}
\right )}} \zeta(\theta_1-\theta_2) ,
\end{equation}
where
\begin{eqnarray}
\zeta(\theta)=c \sinh{\theta/2} \; \exp \left ( \int_0^\infty dk
\frac{
\sin^2 (\frac{k}{2}(\theta/\pi+i))
\sinh (\frac{1-\xi}{2}k)}
{k \sinh \left ( \frac{\xi
k}{2} \right ) \sinh \left (k \right ) 
\cosh{\left ( \frac{k}{2} \right ) }} \right ) \\
c=\left ( \frac{4}{\xi}  \right ) ^{\frac{1}{4}} 
\exp \left ( 
\frac{1}{4} \int_0^\infty 
\frac{
\sinh \left ( \frac{ k}{2} \right )
\sinh \left (\frac{1-\xi}{2} k \right ) }{k \sinh \left ( \frac{\xi
k}{2} \right )  \cosh^2 \left ( \frac{k}{2} \right )}  \right ); \; \;
d=\frac{1}{2 \pi \xi c}.
\end{eqnarray}
The function $\zeta(\theta)$ is analytic in the physical strip $0\leq
{\rm Im} \theta \leq 2\pi$ and satisfies the following identities
\begin{equation}
\zeta(\theta)S_0(\theta)=\zeta(-\theta), \; 
\zeta(\theta-2 \pi i)=\zeta(-\theta)\ .
\label{zetaid}
\end{equation}

The four particle form-factor is far more complicated and can be
represented as \cite{smirnov}
\begin{eqnarray}
\label{4pp}
&&f^j(\theta_1,...,\theta_4)_{--++}
=\frac{4 \pi^3}{\beta}\xi M\sqrt{A} d^2 \prod_{k<l}
\zeta(\theta_k-\theta_l) \nonumber\\
\times&&\prod_{m,n=1,2}\left({\sinh[(\theta_{2+m}-\theta_n-i \pi)/\xi]}
\right)^{-1} \nonumber\\
\times&&2 \sinh[{(\theta_4+\theta_3-\theta_1-\theta_2-2 \pi i)/2 \xi}]
\nonumber\\
\times&&\exp(-\frac{1}{\xi} \sum_k \theta_k)  
\int_{-\infty}^\infty \frac{ d \alpha}{i \pi}\prod_k
\varphi(\alpha-\theta_k) \cosh(\alpha-\frac{1}{2}\sum_k \theta_k) \nonumber\\
\times&&\Delta(e^{2 \alpha/\xi}|e^{2 \theta_1/\xi},
e^{2 \theta_2/\xi}|e^{2 \theta_3/\xi},e^{2 \theta_4/\xi}). 
\end{eqnarray}
Here $\zeta(\theta), c, d$ have been previously defined,  
and
\begin{eqnarray} 
\varphi(\theta)=\xi^\frac{1}{2} \exp \left(-
\int_0^\infty  d k \frac{2\left (\sin^2{\left (\frac{k \theta}{2 \pi}
\right )}
\sinh{\left (\frac{1+\xi}{2}k \right )}
+\sinh^2{ \left (\frac{ k}{4} \right )} \sinh{ \left (
\frac{1-\xi}{2}k \right )} \right )}{k
\sinh{\left (\frac{\xi k}{2} \right )} \sinh{\left (k \right )}} \right)
\nonumber\\
= \xi^\frac{1}{2}\exp 
\left(- \int_0^\infty  d k \frac
{2\left (\sinh^2{ \left (\frac{ k}{4} \right )} \sinh{ \left (
\frac{1-\xi}{2}k \right )} \right )}{k
\sinh{\left (\frac{\xi k}{2} \right )} \sinh{\left (k \right )}}
\right )
\prod_{n=0}^{\infty} \left | 
\frac{\Gamma \left(
\frac{1}{4}+\frac{\xi}{2 } n-i\frac{\theta}{2 \pi}\right ) \Gamma
\left ( \frac{3}{4}+\frac{\xi}{2} (n+1) \right )}
{\Gamma \left ( \frac{3}{4}+\frac{\xi}{2}(n+1)-i\frac{\theta}{2 \pi}\right )
\Gamma \left ( \frac{1}{4}+\frac{\xi}{2 } n \right)}
\right |^2,
\end{eqnarray}
\begin{eqnarray}
&&\Delta(e^{{2\alpha}/{\xi}}|
e^{{2\theta_1}/{\xi}},e^{{2\theta_2}/{\xi}}|
e^{{2\theta_3}/{\xi}},e^{{2\theta_4}/{\xi}})
=\frac{1-e^{2\pi i/\xi}}{8}\bigg[
e^{4\alpha/\xi}(1+e^{2\pi i/\xi})\nonumber\\
&&\qquad-e^{2\alpha/\xi}e^{i\pi/\xi}\sum_{j=1}^4e^{2\theta_j/\xi}
+e^{2(\theta_1+\theta_2+i\pi)/\xi}+e^{2(\theta_3+\theta_4)/\xi}
\bigg].
\end{eqnarray}
The function $\varphi(\theta)$ is even and, as can be easily see from the
last expression, has poles and simple zeros in the physical strip,
$0\leq {\rm Im} \theta \le 2 \pi$, at the points $\theta=i\pi/2+i\pi \xi k$,
$k \geq 0$ and   $\theta=i 3\pi/2 +i\pi \xi k$,
$k \geq 1$. As $\theta \rightarrow \pm\infty$ it behaves like $
\varphi(\theta)\sim 2 \exp (\mp\frac{1}{2} (1+1/\xi)\theta)$.

We note that the above integral expression is valid only in the regime $\xi>2$.
In general it is necessary to regularize the $\alpha$-integral in
(\ref{4pp}) \cite{smirnov}, but for $\xi >2$ no such regularization is
required. 

The form-factors with different orderings of
$\varepsilon_1,...,\varepsilon_4$ 
can be obtained from (\ref{4pp}) using the symmetry property (\ref{order}).
Using (\ref{zetaid}) we obtain
\begin{eqnarray}
&&f^j(\theta_1,\theta_2,\theta_3,\theta_4)_{-,+,-,+} \nonumber\\
&&=\frac{f^j(\theta_1,\theta_3,\theta_2,\theta_4)_{-,-,+,+}
-f^j(\theta_1,\theta_2,\theta_3,\theta_4)_{-,-,+,+}
S^{-+}_{+-}(\theta_2-\theta_3)}{
S_{+-}^{+-}(\theta_2-\theta_3)} \\
&&f^j(\theta_1,\theta_2,\theta_3,\theta_4)_{+,-,-,+}= \nonumber\\
&&\frac{f^j(\theta_2,\theta_3,\theta_1,\theta_4)_{-,-,+,+}-
f^j(\theta_1,\theta_2,\theta_3,\theta_4)_{-,-,+,+} S_{+-}^{-+}
(\theta_1-\theta_3) S_0(\theta_1-\theta_2)}{S_{-+}^{-+}(\theta_1-\theta_2)
S_{+-}^{+-}(\theta_1-\theta_3)} \nonumber\\
&&-\frac{f^j(\theta_1,\theta_3,\theta_2,\theta_4)_{-,-,+,+}-
f^j(\theta_1,\theta_2,\theta_3,\theta_4)_{-,-,+,+}
S_{+-}^{-+}(\theta_2-\theta_3)}{S_{-+}^{-+}(\theta_1-\theta_2)
S_{+-}^{+-}(\theta_2-\theta_3)}S_{+-}^{-+}(\theta_1-\theta_2).
\end{eqnarray}

The remaining orderings appearing in (\ref{4pp}) can be obtained
using the transformation properties of the current form factors under
charge conjugation \cite{smirnov}
\begin{equation}
f_\mu(\theta_1,...,\theta_{2n})_{\epsilon_1,...,\epsilon_{2n}}
=-f_\mu(\theta_1,...,\theta_{2n})_{-\epsilon_1,...,-\epsilon_{2n}}.
\end{equation}
\vskip 1cm
We can now use the above expressions for the formfactors in the
spectral representation for the optical conductivity (\ref{expansion2}).
The two particle contribution is easily obtained by evaluating the
$\delta$-functions in (\ref{expansion2})
\begin{equation}
\sigma_2(\omega)= \frac{2\Theta(\omega-2M)}{{\omega^2}
\sqrt{{\omega}^2-4M^2}}|f(\theta)|^2 \ .
\label{low}
\end{equation}
Here $\Theta(x)$ is the Heaviside function,
\begin{equation}
f(\theta)=f^j(\theta/2,-\theta/2)_{+-}=-f^j(\theta/2,-\theta/2)_{-+}
\end{equation}
and
\begin{equation}
\theta= 2 {\rm arccosh(\tilde{\omega})}\ ,\; \tilde{\omega}=\omega/2M.
\end{equation}
As mentioned before, (\ref{low}) is the full, exact expression for the
optical conductivity for frequencies smaller than $4M$.

After some calculations the four particle contribution can be cast in
the form
\begin{eqnarray} 
\label{four}
&& \sigma_4(\omega)=\frac{\Theta(\omega-4M)}{\omega 
192 \pi^2 M^2} \sum_{\varepsilon_i}
\sum_{\sigma=\pm}
\int_{-a}^{a}
 d\theta \; \int_{-b(\theta)}^{b(\theta)} d \gamma
\nonumber\\
&&\times|f^j(g-\frac{\sigma\alpha}{2},g+\frac{\sigma\alpha}{2},
g+\theta+\gamma,
g-\theta+\gamma)_{\varepsilon_1...\varepsilon_4}|^2
\nonumber\\
&&\times\left\{\left(
\sqrt{\cosh^2{\theta}\sinh^2{\gamma}+\tilde{\omega}^2}
-\cosh{\theta}\cosh{\gamma}
\right)^2-1\right\}^{-\frac{1}{2}}\nonumber\\
&&\times\left[\cosh^2{\theta}
\sinh^2{\gamma}+\tilde{\omega}^2\right]^{-\frac{1}{2}}
\ ,
\end{eqnarray}
where 
\begin{eqnarray}
a&=&{\rm arccosh(\tilde{\omega}-1)}, \ 
b(\theta)={\rm arccosh\left[\frac{\tilde{\omega}^2-1-\cosh^2{\theta}}
{2 \cosh{\theta}}\right]}, \nonumber\\
g&=&\ln\left[\frac{\cosh(\alpha/2)+\exp(-\gamma)\cosh\theta}
{\tilde{\omega}}\right], \nonumber\\
\alpha&=&2 {\rm arccosh}\left[
\sqrt{\cosh^2{\theta}\sinh^2{\gamma}+\tilde{\omega}^2}- 
\cosh{\theta}\cosh{\gamma}\right]. \nonumber 
\end{eqnarray}
The remaining integrals in (\ref{four}) as well as the function
$\zeta(\theta)$ have to be evaluated numerically. The latter is easily
done to very high precision. The multiple integrals in the expression
for the four-particle contribution are much more difficult to evaluate
numerically. We estimate the precision of our results to be of the
order of $10^{-4}$.

\begin{figure}[ht]
\begin{center}
\noindent
\epsfxsize=0.5\textwidth
\epsfbox{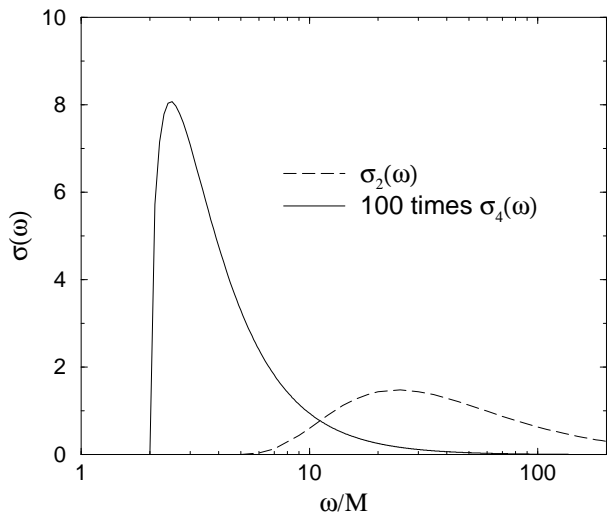}
\end{center}
\caption{\label{fig:ff}
Two particle (solid line) versus one hundred times the four particle
contribution (dashed line) in the form factor expansion for $\beta^2=0.9$.
}
\end{figure}
We have evaluated the two and four particle contributions to the
optical conductivity for several values of $\beta$.
The two and one hundred times the four-particle contributions
for $\beta^2=0.9$ are presented in Fig.1. Most importantly, the square
root singularity, being a characteristic feature of band insulators,
is suppressed by the momentum dependence of the soliton-antisoliton
form factor and reappears {\sl only} at the LE point
$\beta^2 =1/2$ (the behaviour of the optical conductivity in the
vicinity of the LE point is shown in Fig. \ref{thres} and
discussed in the next section).

The rounding off of the singularity is due to the $\theta$ dependence
of the term (\ref{low}). A similar behavior was previously noted for
the Hubbard model at half-filling \cite{hubb} which corresponds to the
special SU(2)-symmetric point $\beta^2 = 1$. We find that for {\sl
any}\footnote{Note that in order to determine behaviour just above the
threshold we need to consider only the two-particle contribution to
the optical conductivity.} $\beta^2\neq 1/2$ there is a square root
``shoulder'' $\sigma(\omega)\propto\sqrt{\omega-\Delta}$ for
$\omega/\Delta-1\ll 1$.

The four particle contribution to $\sigma(\omega)$ is seen to be
insignificant at low energies and becomes larger than the two particle
contribution only at $\omega \approx 180 M$ for $\beta^2=0.9$.
This suggests that the optical conductivity is well described by the 
combination of 2 and 4-particle contributions up to several hundred 
times the mass gap.
Computation of higher order terms in (\ref{expansion2}) becomes
cumbersome and probably of no physical interest, since the previous
analysis suggests that they become important outside the region of
applicability of the field theory approach to physical systems
\footnote{The field theory approximation is appropriate as long as the
frequencies considered are much less than the band width, which is
the UV scale in the problem. In practical applications one is unlikely
to encounter a situation where the band width is more than 1000 times
the spectral gap.}.
Nevertheless it is interesting from a theoretical point of view to
determine their importance at very high frequencies, which we will
do using a different approach in section \ref{sec:largeen}.

\section{Vicinity of the Luther-Emery point}

As we have indicated in the previous section, the LE point is quite
special. Let us now discuss the behaviour of the optical conductivity
in the vicinity of the LE point. This will exemplify some differences
between Mott insulators and conventional band insulators. 

Let us recall that the SGM is equivalent to the Massive Thirring Model
(MTM) \cite{SG1} 
\begin{equation}
\label{mtm-action}
{\cal S}_{MTM}=\int \; d^2 x  \left [ i \bar\psi \gamma^\mu \partial_\mu
\psi-\frac{g}{2}\left (\bar\psi \gamma_\mu \psi \right )^2-m\bar\psi
\psi \right ]
\end{equation}
with the following identifications
\begin{eqnarray}
\label{eq:equivalence}
j^\mu=\bar 
\psi \gamma^\mu\psi&=&-\frac{\beta}{2 \pi} \epsilon^{\mu \nu} \partial_\nu \phi\nonumber\\
\frac{g}{\pi}&=&\frac{1}{2}\left ( \frac{1-\xi}{\xi} \right ) \; .
\end{eqnarray}

Here $\psi=\pmatrix{\psi_1\cr \psi_2}$ and $\bar \psi=\psi^\dagger\gamma^0$ are the usual
two-component Fermi fields and the gamma matrices are chosen as
\begin{equation} 
\gamma^0=\sigma^1, \; \gamma^1=i\sigma^2, \nonumber
\end{equation}
where $\sigma^i$ are Pauli matrices. The fermions in the MTM
correspond to solitons and antisolitons in the SGM. From
(\ref{eq:equivalence}) it follows that for $\xi=1$ ($\beta^2=1/2$) the
fermions become non interacting and the action (\ref{mtm-action})
describes a two-band system of free, spinless fermions. 
In the absence of doping, in the ground state the upper band is empty
while the lower one is completely filled. Thus the LE point describes
a conventional band insulator. In fact, at $\xi=1$ Eq. (\ref{low})
yields a square root singularity above the threshold, in agreement
with this interpretation.

\begin{figure}[ht]
\begin{center}
\noindent
\epsfxsize=0.5\textwidth
\epsfbox{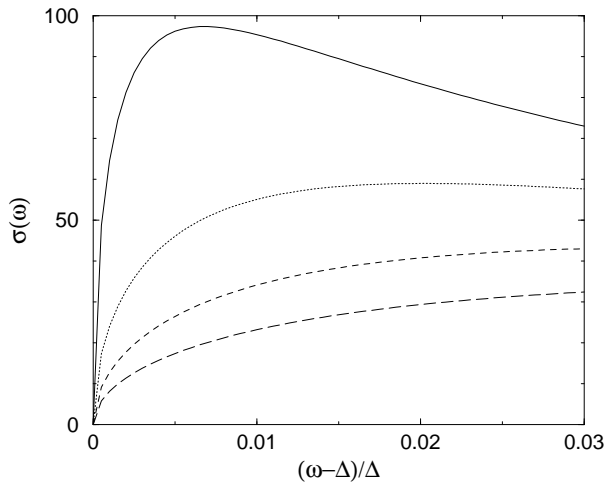}
\end{center}
\caption{\label{thres}
Threshold behaviour of the optical conductivity close to the
Luther-Emergy point for four different values of $\beta$; $\beta=0.72$
(solid), $\beta=0.73$ (dotted), $\beta=0.74$ (dashed) and $\beta=0.75$
(long dashed) .
}
\end{figure}

However, as soon as we deviate from the LE point this singularity
immediately disappears and the optical conductivity {\sl vanishes} at
the optical gap. Close to the LE point (\ref{low}) implies the following
analytical expression valid for ${\tilde\omega} - 1 \ll 1$
\begin{eqnarray}
\sigma(\omega) \propto \frac{\sqrt{\tilde\omega^2 - 1}}{[\tilde\omega^2
-1] + \xi^2\sin^2\gamma},~~ \gamma = \pi\left(\frac{1}{2\beta^2} -
1\right).
\end{eqnarray} 
The threshold behaviour of $\sigma(\omega)$ for several values of
$\beta$ is shown in Fig. \ref{thres}.

We see that the square root singularity above $\omega=\Delta$ for
$\beta^2=1/2$ is replaced by a maximum occurring at
$\omega_{max}=\Delta +{\cal O}(\gamma^2)$.
As we take $\beta^2\to 1/2$ from above, $\omega_{max}$ approaches
$\Delta$ and $\sigma(\omega_{max})$ diverges.

\section{Large energy behavior}
\label{sec:largeen}

In order to obtain a rough estimate of the importance of the
contributions involving six or more particles in the FF sum
and to obtain a complete picture of the optical conductivity, we now
calculate $\sigma(\omega)$ at large energies. 
A convenient method to do this is Conformal Perturbation Theory
(CPT)\cite{cptZ1,cptZ2}. Viewing the SGM as a Gaussian model
\begin{equation}
S_{Gauss}=\int \; d^2 x \frac{1}{16\pi}(\partial_\nu
\phi )^2
\end{equation}
perturbed by the relevant operator $2 \cos(\beta \phi)$, we can
formally obtain correlation functions of the perturbed theory by
\begin{equation}
\label{corr.cpt}
\langle{\cal O}({\bf x_1}){\cal O}({\bf x_2})...{\cal O}({\bf x_n}) \rangle=
\frac{\langle{\cal O}({\bf x_1}){\cal O}({\bf x_2})...{\cal O}({\bf x_n})
e^{-2\mu \int \;  d^2 x \cos(\beta \phi)} \rangle_{CFT}}{\langle e^{-2\mu \int \; d^2 x
\cos(\beta \phi)} \rangle_{CFT}} 
\end{equation}
We use a normalization in which
\begin{equation}
\label{normalization}
\langle e^{i\beta \phi({\bf x_1})} \ldots e^{i\beta \phi({\bf
x_n})}e^{-i\beta \phi({\bf y_1})}
\ldots e^{-i\beta \phi({\bf y_n})} \rangle_{CFT}=
\frac{\prod_{i <j}^n|{\bf x_i}-{\bf x_j}|^{4 \beta^2}|{\bf y_i}-{\bf
y_j}|^{4 \beta^2}}{\prod_{i,j=1}^n |{\bf x_i-y_j}|^{4 \beta^2}}\ .
\end{equation}
The two-point function of electric currents can for example be
obtained by considering
\begin{equation}
\langle j({\bf x})\ j({\bf x^\prime})\rangle= A
\langle\partial_\tau \phi(x,\tau) \partial_{\tau '} \phi(x',\tau ') \rangle=
A\lim_{\alpha \rightarrow 0 } 
\frac{1}{\alpha^2}\partial_\tau
\partial_{\tau '}\langle e^{i\alpha \phi(x,\tau)}e^{-i\alpha
\phi(x',\tau ')} \rangle .
\label{derivates}
\end{equation}
Using (\ref{corr.cpt}) we have
\begin{eqnarray}
&&\langle j({\bf x}) j({\bf x^\prime})
\rangle\sim A\lim_{\alpha \rightarrow 0 } 
\frac{1}{\alpha^2}\partial_\tau
\partial_{\tau '}\langle e^{i\alpha \phi({\bf x})}e^{-i\alpha \phi({\bf x^\prime})} \rangle\nonumber\\
&=&A\frac{\sum_n \frac{1}{n!}(\frac{-\mu}{2})^n\int \;
d^2\omega_1 ... d^2\omega_n \sum_{l_k=\pm1}
\lim_{\alpha \rightarrow 0 } 
\frac{1}{\alpha^2}\partial_\tau
\partial_{\tau '}\langle e^{i l_1 \beta \phi({\bf \omega_1})}...e^{i l_n \beta
\phi({\bf \omega_n})}
e^{i\alpha \phi({\bf x})}e^{-i\alpha \phi({\bf x^\prime})} \rangle_{CFT}}
{\sum_n \frac{1}{n!}(\frac{-\mu}{2})^n\int \;
d^2\omega_1 ... d^2\omega_n \sum_{l_k=\pm1}
\langle e^{i l_1\beta \phi({\bf \omega_1})}...e^{i l_n \beta \phi({\bf
\omega_n})}
\rangle_{CFT}}.\nonumber\\
\label{jj.cpt}
\end{eqnarray}

Eq.(\ref{jj.cpt}) provides a perturbative expansion in integer powers
of the scale $\mu$. Using the exact solution of the SGM it is possible
to relate $\mu$ to the physical soliton mass $M$ \cite{abz} 
\begin{equation}
\mu=\frac{\Gamma(\beta^2)}{\pi \Gamma(1-\beta^2)}\left [
M\frac{\sqrt{\pi}
\Gamma(1/2+\xi/2)}{2 \Gamma(\xi/2)} \right ]^{2-2 \beta ^2}.
\label{gap}
\end{equation}

In our case the CPT expansion is free of ultraviolet divergences 
(which would cause further complications \cite{cptZ2,cds,fms})
as long as $\xi<\infty$ but is known to suffer from infrared divergences.
For example, the correlation function of bosonic exponents in the
first line of (\ref{jj.cpt}) develops infrared divergences at order
$O(\mu^{2n})$ for $\beta^2\leq 1-1/2n$ \cite{leclair}. 
Here we only consider the term of second order in $\mu$, which is free of
divergences as long as $\beta^2>1/2$. 
For more general calculations one could follow the approach suggested
in \cite{cptZ2} (see also \cite{cpt1}).
 
\begin{figure}[ht]
\begin{center}
\noindent
\epsfxsize=0.6\textwidth
\epsfbox{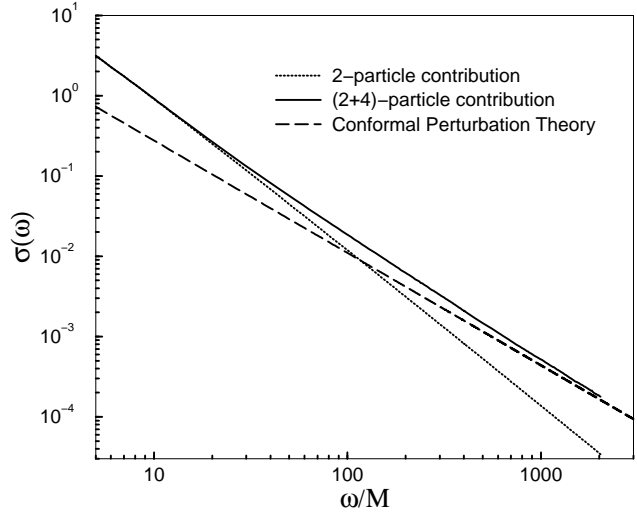}
\end{center}
\caption{\label{fig:cpt}
Comparison between the 2+4-particle contribution to the optical conductivity
for $\beta^2=0.9$ and the results from bare perturbation theory.
}
\end{figure}

We can now calculate the current-current correlation function to
leading order in CPT and use the result to determine the optical
conductivity at large energies. The $\omega_j$-integrals in
(\ref{jj.cpt}) are carried out using the methods of
\cite{integrals}. We find
\begin{eqnarray}
&&\sigma(\omega)=A 2^{9-4\beta^2}\left(\frac{\pi^2\beta}
{\Gamma(2\beta^2)}\right)^2\mu^2 \omega^{(4 \beta^2-5)}\nonumber\\
&&=\frac{8\pi^3\beta^2 A}{\omega\Gamma^2(1-\beta^2)\Gamma^2(\frac{1}{2}+\beta^2)}
\left[\frac{\Gamma(\frac{\xi}{2})}{2\sqrt{\pi}\Gamma(\frac{1+\xi}{2})}
\frac{\omega}{M}\right]^{4\beta^2-4}.
\label{high}
\end{eqnarray}
We emphasize that the ratio of the coefficients of the high- and
low-energy asymptotics (\ref{high}), (\ref{low}) is {\sl fixed}
\cite{smirnov},\cite{lukyanov}. In other words, the amplitude of the
power law in (\ref{high}) is tied to the overall factor in (\ref{low}) and
the form factor expansion must approach the perturbative result in the
large-$\omega$ limit. 
A comparison between the form factor results and (\ref{high}) is shown in
Fig.\ref{fig:cpt}. 
We see that the asymptotic regime is not yet reached at energies as
high as $\omega \sim 1000 M$. In practical terms this implies that
perturbation theory (PT) cannot be used to make contact with experiment.
We note that the contributions due to intermediate states with
6,8... particles are all positive and will make the agreement of the
form factor sum with PT in the region  $\omega\approx
1000M$ only worse.

A good way to overcome these deficiencies of bare PT is to carry out a
renormalization-group (RG) improvement as performed in \cite{gia1}. 
Following \cite{abz} we describe the SGM as the $\bar {\cal J} {\cal J}$
perturbation of the $SU(2)_1$ Wess-Zumino-Witten model
\begin{equation}
S_g=S_{SU(2)_1}+\frac{g_\parallel}{2\pi} \int \ d^2 x  \; \bar {\cal
J}_0 
{\cal J}_0+
\frac{g_\perp}{4 \pi} \int \ d ^2 x \; (\bar {\cal J}_+ {\cal J}_-+ 
\bar {\cal J}_- {\cal J}_+).
\end{equation}
Here ${\cal J}_\alpha$ and $\bar {\cal J}_\alpha$ 
are the left and right currents of
SU(2)$_1$, normalized by the following operator product expansions
\begin{eqnarray}
\nonumber
{\cal J}_0(z){\cal J}_0(0)=\frac{1}{2z^2}+O(1) \\
{\cal J}_0(z){\cal J}_\pm(0)=\pm\frac{1}{z}{\cal J}_\pm(0)+O(1) 
\\ \nonumber
{\cal J}_+(z){\cal J}_-(z)=\frac{1}{z^2}+\frac{2}{z}{\cal J}_0(0)+O(1). 
\end{eqnarray}
The SU(2)-symmetric perturbation $g_\perp =g_\parallel$ corresponds to
$\xi \rightarrow \infty$. 
The RG equations for the Sine-Gordon model can be cast in the form \cite{abz}
\begin{equation}
\frac{dg_\perp}{dt}=\frac{g_\parallel g_\perp}{1+\frac{g_\parallel}{2}}\ ,\quad
\frac{dg_\parallel}{dt}=\frac{g^2_\perp}{1+\frac{g_\parallel}{2}}\ ,
\label{rgeq}
\end{equation}
where $t$ is the RG scale. The solution of (\ref{rgeq}) is \cite{abz}
\begin{equation}
g_\perp=4\frac{1-\beta^2}{\beta^2}\frac{\sqrt{q}}{1-q}\ ,\quad
g_\parallel=2\frac{1-\beta^2}{\beta^2}\frac{1+q}{1-q}\ ,
\label{rg1}
\end{equation}
where
\begin{equation}
q\left(\frac{(1-q)\beta^2}{4(1-\beta^2)}\right)^{2\beta^2-2}=e^{(4-4\beta^2)
(t-t_0)}.
\label{rg2}
\end{equation}

\begin{figure}[ht]
\begin{center}
\epsfxsize=0.5\textwidth
\epsfbox{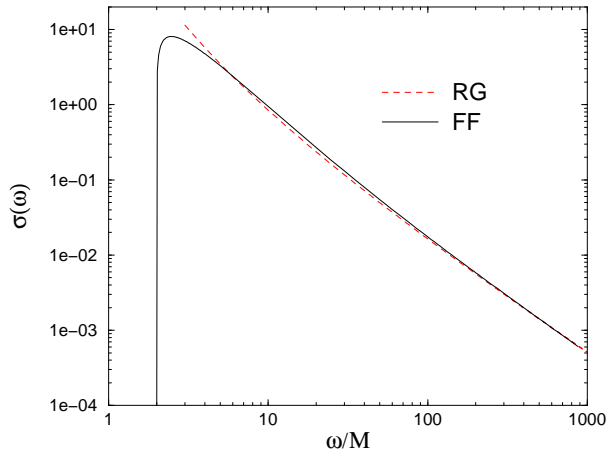}
\end{center}
\caption{\label{rg.fig}
Comparison between the 2 and 2+4-particle contribution to the optical
conductivity for $\beta^2=0.9$ and the RG improved PT.
}
\end{figure}

The perturbative result (\ref{high}) is expressed in terms of
$g_\parallel$, $g_\perp$ noting that
\begin{equation}
g_\parallel^2-g_\perp^2=\frac{4}{\xi^2}\ ,
\end{equation}
and by considering $q\to 0$ in (\ref{rg1}), (\ref{rg2}). Finally one
may fix the scale $t_0$ by simply choosing it the same as in \cite{abz}
$t-t_0=\ln\left(\frac{\sqrt{\pi}e^{3/4}M}{2^{3/2}\omega}\right)$ 
\footnote{In order to choose $t_0$ is a meaningful way one needs to
calculate the first subleading term in the CPT expansion, which is
outside the scope of these proceedings.}
and obtain modulo terms of higher order in the couplings
\begin{equation}
\sigma(\omega)=\frac{\pi^3\beta^6
g_\perp^2}{2\omega\Gamma^2(2-\beta^2)\Gamma^2(\frac{1}{2}+\beta^2)} 
\left[\frac{\Gamma(\frac{\xi}{2})e^{3/4}\sqrt{\xi}}{2^{7/2}\Gamma(\frac{1+\xi}{2})}
\right]^{4\beta^2-4}.
\label{rg}
\end{equation}

The RG improved result (\ref{rg}) for $\sigma(\omega)$ is compared to the
form factor result (or more precisely the sum of the two and
four-particle contributions) in Fig.\ref{rg.fig}. We see that on the
level of accuracy of a log-log plot the agreement is rather good down to
energies of the order of $5M$. Combining the FF results with the RG
improved perturbation theory we then obtain a good description
of the optical conductivity over the whole frequency range.
We believe that the good agreement of RG with the exact result even at
rather small frequencies is probably a particular feature of the
correlation function considered here. For other correlators like for
example the spectral function there is no reason to believe that RG
will work as well as it does for the optical conductivity.

\section{Applications}

Let us now turn to applications of our results for the optical
conductivity. There are several materials which are believed to be
one-dimensional Mott insulators in one of their phases.
These are quasi-1D antiferromagnets like ${\rm KCuF_3}$ \cite{AFM},
Carbon nanutubes \cite{nanotubes}, possibly the striped phase in
${\rm La_{1.67}Sr_{0.33}NiO_4}$\cite{stripes} and organic conductors
\cite{exp,bj}. 
Here we will concentrate on the latter due to the availability of
extensive optical data. Of particular interest for our purposes are
the ${\rm (TMTTF)_2X}$ and ${\rm (TMTSF)_2X}$ families, where X is an
inorganic monoanion. These materials exhibit a rich phase diagram as a
function of temperature and pressure \cite{bj} and at sufficiently
high temperatures are believed to be 1D Mott insulators.
The ${\rm (TMTTF)_2X}$ family are presumably good examples of 1D Mott
insulators, but as the optical gap in these materials is of the order
of the bandwidth a field theory description is inappropriate.

For the $\rm (TMTSF)_2X$ Bechgaard salts the ratio of optical gap to
bandwidth is small and a field theory description is possible.
The Bechgaard salts are highly anisotropic materials and can be
modelled as weakly coupled, quarter-filled chains. 
It was suggested in \cite{giamarchi} that at energies or temperatures
sufficiently far above the 1D-3D crossover scale $E_{\rm cr}$, the 
interchain coupling becomes ineffective and a description in terms of
a purely 1D model with charge sector (\ref{sg}) should be possible.
This is a very nontrivial assertion as the microscopic lattice
Hamiltonian appropriate for these systems is not given by a simple
extended Hubbard model like (\ref{Hext}), but includes an explicit
dimerization \cite{pm}. The low-energy effective field theory obtained by
bosonization is therefore not given by the simple form (\ref{Hext2}),
but contains other perturbing operators as well.
At present there is also some uncertainty regarding the value of 
$E_{\rm cr}$ because interactions can renormalize its bare value, set
by the interchain coupling, downwards \cite{boies}. 

\begin{figure}[ht]
\begin{center}
\epsfxsize=0.6\textwidth
\epsfbox{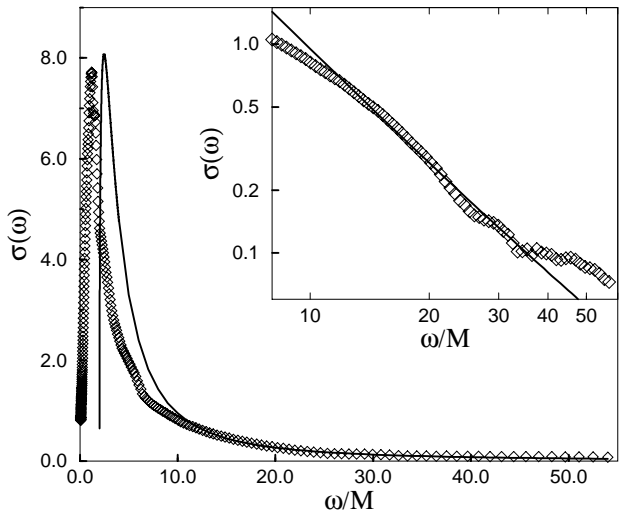}
\end{center}
\caption{
\label{fig:exp}
Comparison between the optical conductivity  calculated in the
SGM for $\beta^2=0.9$ (solid lines) and measured optical
conductivity for $\rm (TMTSF)_2PF_6$ from Ref.\cite{exp} (diamonds). The
inset shows the same comparison on a logarithmic scale.}
\end{figure}

There is a lot of ambiguity in fitting our results to the data. The
value of the optical gap $2M$ is not known and, as discussed above, we
cannot calculate the overall normalization of $\sigma(\omega)$. We
therefore use these as parameters in order to obtain a good fit at
large $\omega$ (where the theory is expected to work best as 3D
effects are unimportant) to the data \cite{exp} for any given value of
$\beta$. We obtain reasonable agreement
with the data for $\beta^2\approx 0.9$, which corresponds to a Luttinger 
liquid parameter of $K_\rho=\beta^2/4\approx 0.23$. This value is 
consistent with previous estimates (see the discussion in \cite{exp}). 

As is clear from Fig.\ref{fig:exp}, the model (\ref{sg}) seems to apply
well at high energies, but becomes inadequate at energies of the order
of about $10$ times the Mott gap ($\approx 1600/{\rm cm}$ in $\rm
(TMTSF)_2PF_6$). Spectral weight is transferred to lower energies and
physics beyond that of a pure 1D Mott insulator emerges. 

There are at least two mechanisms that should be taken into account in
this range of energies. Firstly, as mentioned before, a small
dimerization occurs in the 1D chains and will almost certainly affect
the structure of $\sigma(\omega)$ around its maximum. Secondly, the
interchain hopping is no longer negligible \cite{interchain} and ought
to be taken into account.

\vskip 1cm
{\bf Acknowledgments:}

We are grateful to A. Schwartz for generously providing us with the
experimental data. We have benefitted greatly from discussions with
F. Gebhard, T. Giamarchi and E. Jeckelmann. We are especially grateful
to S. Lukyanov for suggesting the RG analysis to us and for important
comments and discussions. We thank the Isaac Newton Institute for
Mathematical Sciences, where this work was completed, for hospitality.
F.H.L.E. is supported by the EPSRC under grants AF/100201 and
GR/N19359.

\end{document}